\documentclass{llncs}
\usepackage{graphicx}
\usepackage{float}
\graphicspath{ {.} }
\begin{document}

\title{Applying Data Compression Techniques on Systolic Neural Network Accelerator}
\titlerunning{Neural Network Accelerators}  
%
\author{Navid Mirnouri}
\authorrunning{Navid Mirnouri} 
%
\tocauthor{Navid Mirnouri}
\institute{
Amirkabir University of Technology, Hafez St., Tehran, Iran\\
\email{n72m@aut.ac.ir}
\\[12pt]
\textbf{Technical Report}\\
September 2016}

\maketitle              

\begin{abstract}
New directions in computing and algorithms has lead to some new applications that have tolerance to imprecision. Although, These applications are creating large volumes of data which exceeds the capability of today's computing systems. Therefore, researchers are trying to find new techniques to alleviate this crisis. Approximate Computing is one promising technique that uses a trade off between precision and efficiency of computing. Acceleration is another solution that uses specialized logics in order to do computations in a way that is more power efficient. Another technique is Data compression which is used in memory systems in order to save capacity and bandwidth.\\
\end{abstract}

\section{Introduction}
\begin{figure}
	\centering
  \includegraphics[width=2.5cm]{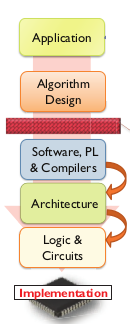}
  \caption{Layers of approximate computing[2].}
  \label{fig:layers}
\end{figure}

Emerging applications in mining, learning and big data have created a large workload for computing systems. While technology improvement is diminishing, designers are trying to find alternative solutions for this crisis\cite{Moreau:Wyse}. Approximate Computing is a promising solution, which utilize resilience of application in order to gain efficiency in all layers of computing stack\cite{Venkataramani:Chakradhar}. These techniques can be applied in each layer of computing: application, architecture and circuit design. Figure\ref{fig:layers} illustrate this hierarchical architecture.
Specialized logic including accelerators and programmable logic is another solution\cite{Venkataramani:Chakradhar}. In some previous works, neural networks were used as an accelerator in order to be an alternative for certain part of code\cite{Esmaeilzadeh:Sampson}. SNNAP is a neural accelerator which uses FPGA as an accelerator for implementation of Neural Processing Units (NPUs)\cite{Moreau:Wyse}. SNNAP has been implemented on Zynque board which is a Field Programmable logic (FPGA). SNNAP uses ACP port for communication between NPUs and the processor. We believe that by applying some techniques including data compression, effective memory bandwidth could be increased.
Linearly Compressed Page(LCP)\cite{Pekhimenko} is one solution, which uses two compression techniques: Based Delta Immediate Compression\cite{Pekhimenko : Seshadri} and Frequent Pattern Compression\cite{Alameldeen}.
In future section we give a short review on approximate computing in different layer. Then we investigate SNNAP, after that we study Data compression techniques including: Based Data Immediate, Frequent Pattern and Linearly Compressed Page.

\section{Approximate Computing}
 In today computing systems, energy efficiency is a key concern. Some application like mining, image processing and learning have tolerance to imprecision. Designing a platform for approximate computing is a trade off between energy-efficiency and precision\cite{Han}.
At circuit design level there some approaches like time scaling or voltage scaling have been proposed\cite{Chippa}.
Different approaches at architectural level could be categorized in three groups\cite{Han}: application specific accelerators, programmable processors and approximate memories. In \cite{Venkataramani} a programmable processor has been proposed which uses ISA extension to use some ISA for specifying the accuracy of processing.
Approximation in software level can reduce the run-time of application, it could be achieved via different approaches\cite{Venkataramani}. Specifying the part of the code to be executed approximately is an important challenge\cite{Venkataramani}.

\section{Neural Networks}

Neural network accelerators has been proposed in previous works\cite{Esmaeilzadeh:Sampson}\cite{Chen}\cite{Grigorian}. The key idea is to specify part of the code that have the tolerance to imprecision and transform them to Neural Networks. Implementation of neural networks can be both in hardware\cite{Esmaeilzadeh:Sampson}\cite{Merolla}\cite{Seo} or software\cite{Chen}\cite{Grigorian}.
\subsection{ Systolic Neural Network accelerator in Programmable logics[3]}

SNNAP is a neural network accelerator which uses programmable system on chips (PSocs) to implement Neural Processing units. PsoCs are heterogeneous architectures which have both hard processor and programmable logics on the same die\cite{Esmaeilzadeh:Sampson}. Integration of the CPU and programmable logic can be beneficial for interconnections' bandwidth comparing to traditional FPGAs.

Some important challenges in SNNAP should be mentioned before going through more details.\\
1. Neural processing unit should be implemented on FPGA in a way that uses resources to minimize energy consumption.\\
2. Communication between NPU and processor should be low latency in order to be efficient for those applications which have smaller part of approximate code. Processing batches of requests is one solution.\\
3. In order for the system to be more energy efficient, Processor and NPU should be able to work independently. For example, when NPU is working CPU can be hibernated with special commands.\\
4. After transformation,different applications have different Nns. This variation in neural network topology should be handled appropriately without much of hardware effort and reprogramming the FPGA.\\

SNNAP invokes MLP which is a multi-layer directed graph\cite{Esmaeilzadeh:Sampson}. For implementation of MLP, SNNAP uses Digital Signal Processing(DSP) slices of FPGA.

\begin{figure}
	\centering
  \includegraphics[width=6cm]{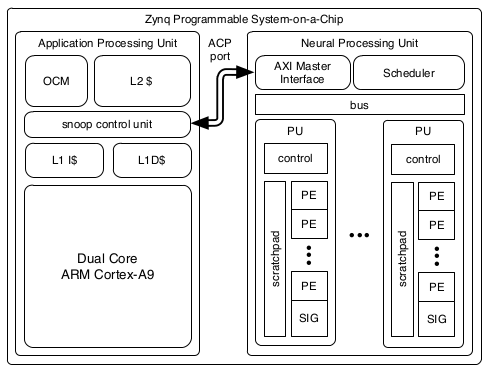}
  \caption{Architectrual Design of SNNAP\cite{Esmaeilzadeh:Sampson}.}
  \label{fig:arch}
\end{figure}
\section{Data Compression}
Data compression is an approach to reduce capacity and bandwidth requirements by recognizing redundancy of data in applications \cite{Yazdanb}. These techniques can be applied in order to improve cache performance, memory bandwidth or interconnections. In this section we are going to study some of these techniques.

Base-delta-immediate is a compression technique for increasing on-chip cache's capacity. Data in cache lines are not wide in range. The key idea in BDI is to use low dynamic range of data in cache line and represent data with a base value and an array of deltas\cite {Pekhimenko : Seshadri}.
Frequent Pattern Compression (FPC) realize that large volumes of records in memory are occurring frequently. This technique tries to compress these record with fewer bits \cite{Pekhimenko}.
Linearly Compressed Page (LCP) is a technique which uses two previous algorithm to gain more efficiency than prior works \cite{Pekhimenko}. The key idea in LCP is using a fixed size for each compressed block in order to eliminate complexity of calculating the location of each block in memory. Prior works, used variable-sized of cache line, and this approach can cause a lot of long-latency and complex main memory address calculations\cite{Pekhimenko}. 

\section{Future Works}
In this report some hot topics in computer architecture has been studied briefly. Approximate Computing is a technique that tries to benefit from imprecision tolerance in some applications like mining, learning in order to do computing in a more efficient way. This technique can be applied in different layers of abstraction from application to circuit design.

Neural Acceleration is another direction that tries to invoke Neural Networks in either software or hardware form to improve the efficiency of computing. It selects certain part of the code and transform them to Neural Network. Implementation of Neural networks can be in hardware \cite{Esmaeilzadeh:Sampson}\cite{Merolla}\cite{Seo}or software \cite{Chen}\cite{Grigorian}.

We believe that SNNAP have some potentials for future works. SNNAP is using multi-layered perceptron (MLP), we believe that it can be implemented more efficiently if we use certain classes of Neural networks for different category of applications. There has been a study on implementation of PARSEC and accelerating it with Neural Networks\cite{Chen}. BenchNN uses different kinds of Neural Networks for each benchmark in PARSEC. For example for canneal benchmark, BenchNN uses Hopfield Neural Network (HNN) instead of MLP\cite{Chen}. We believe that future works can address the possibility of customized NPU implementation in SNNAP. 

Another direction is to study data compression techniques that can be applied to SNNAP for improving memory bandwidth. Some of them have been mentioned briefly in this paper, but there are more research opportunities to find a practical data compression that can be applied to SNNAP.  

%
%

\end{document}